\documentclass{elsart}

\usepackage{amssymb}
\usepackage{graphicx}
\newcommand{\bea}{\begin{eqnarray}}
\newcommand{\ena}{\end{eqnarray}}

\renewcommand{\a}{\alpha}
\renewcommand{\b}{\beta}

\begin{document}

\begin{frontmatter}



\title{Prediction of $|V_{13}|$ and the Dirac phase 
in the Neutrino Mixing Matrix$^\star$}
\thanks{a talk given at PANIC02, Osaka, September, 2002}

\author{Takahiro Miura, Tetsuo Shindou and 
Eiichi Takasugi}
\address{Department of Physics,
Osaka University \\ Toyonaka, Osaka 560-0043, Japan}
\begin{abstract}
We considered a model where $|V_{13}|=0$ so that $\delta=0$ at $M_R$ 
and analyzed mixing angles and CP violation angles at the low energy 
scale by using the renormalization group in the MSSM model. In this 
model, parameters at $M_R$ scale are two mixing angles ($\theta_{12}$
and 
$\theta_{23}$), three neutrino masses, and two Majorana phases.  
We found that (1) $\sin^2 2\theta_{12}$ at the low energy scale 
must be less than that at $M_R$ scale, (2) the induced $|V_{13}|$ 
can be as large as 0.05 if masses of neutrinos are around 0.05 eV. 
\end{abstract}

\end{frontmatter}
The recent experimental data by SuperKamiokande[1,2] 
and SNO\cite{SNO-Solar} 
show that the solar neutrino mixing angle is 
$\tan^2 \theta_\odot\simeq 0.34$ and
the atmospheric neutrino mixing angle is $\sin^22\theta_{\rm atm}\simeq 1$, 
while $|\sin\theta_{13}|$=$|V_{13}|<0.16$. These data show a particular 
pattern, that is, mixing angles $\theta_{12}=\theta_\odot$ and 
$\theta_{23}=\theta_{\rm atm}$ are large, while $\theta_{13}$ is 
quite small. \\
There must be some reason why $\theta_{13}$ is so small in comparison 
with the other. By taking this fact seriously, we assume that 
$\theta_{13}=0$ so that the Dirac phase, $\delta$ is zero 
at $M_R$ where neutrino mass matrix is generated. 
The neutrino mixing matrix at $M_R$ 
is described by two mixing angles, three neutrino masses 
and two Majorana 
phases\cite{Bilenky:1980cx,DBD-Majo}. 
In this scenario, the angle $\sin\theta_{13}=|V_{13}|$ and $\delta$ are 
induced by the renormalization group. We construct a model along 
this scenario and observe what is the role of Majorana phases 
for the renormalization group, in what situation the experimentally 
measurable size of $|V_{13}|$ is induced and what is the prediction 
for $\delta$ following the recent work by Miura, Shindou and Takasugi
\cite{MST}. 
\\
{\it The mass matrix at $M_R$}: 
The mass matrix which gives $V_{13}=0$ is generally expressed by 
$m_\nu(M_R) =O {\rm diag}(M_1, M_2e^{i\alpha_0}, M_3e^{i\beta_0}) O^T$, 
where $\alpha_0$ and $\beta_0$ are Majorana phases. $O$ is 
the mixing matrix at $M_R$ scale and is given by 
$O=O_{23}(\theta_{23})O_{12}(\theta_{12})$\\
{\it The neutrino mass matrix at $m_Z$}: 
The mass matrix is given by\cite{RGE-formula}
$m_\nu(m_Z)={\rm diag}(1,1,\a) O D_\nu O^T{\rm diag}(1,1,\a)$, 
where $\a$ is defined by 
$\a \equiv 1- \epsilon =1/\sqrt{I_{\tau}}
= ( \frac{m_Z}{M_R} )^{\frac{1}{8\pi^2}(1+\tan^2\b)(m_{\tau}/v)^2} <1$. 
Here, $v^2=v^2_u +v^2_d$ and $\tan\b = v_u/v_d$ with
$v_i=\langle H_i\rangle(i=u,d)$ for the MSSM model. 
By taking  
$M_{\rm R}=10^{13}({\rm GeV})$ and $2< \tan \beta <50$, we find 
$8\times 10^{-4}< \epsilon < 5\times 10^{-2}$.\\
The followings are known\cite{RGE-formula,Stability,RGEwithMaj}: 
(a) The mixing angles are stable for the case of the hierarchical 
or the anti-hierarchical neutrino mass scheme, $m_1\ll m_2\ll m_3$ 
or $m_3\ll m_1\ll m_2$. (b) The instability occurs for $m_1\simeq m_2$. 
(c) The Majorana phases in neutrino masses may play important 
role\cite{RGEwithMaj}.\\
We focus on the unstable case, 
$M_1\simeq M_2\;, \;0<\Delta_{21}\ll |\Delta_{31}|$, 
$0<\Delta_{21} \ll M_1^2\;,
\;\epsilon M_1^2\simeq \epsilon M_2^2\ll|\Delta_{31}|$, 
where $\Delta_{21}=M_2^2-M_1^2$, $\Delta_{31}=M_3^2-M_1^2$. 
The diagonalization of the neutrino mass matrix is made 
analytically. \\
We parameterize the mixing matrix at $m_Z$ scale as 
\bea
V=\pmatrix{
c_\odot &-s_\odot &-|V_{13}|e^{-i\delta}\cr
s_\odot c_{\rm atm}&c_\odot c_{\rm atm}&-s_{\rm atm}\cr
s_\odot s_{\rm atm}&c_\odot s_{\rm atm}&c_{\rm atm}\cr}
\pmatrix{
1&&\cr
&e^{i\a_M}&\cr
&&e^{i\b_M}\cr}\;,
\ena
and masses as $m_i$. 
Among them, stable quantities are $m_i$ and 
$\Delta m_{\rm atm}^2 =m_3^2-m_2^2$ and $\theta_{\rm atm}$, 
that is, $m_i \simeq M_i$, $\Delta m_{\rm atm}^2 \simeq M_3^2-M_1^2$ 
and $\theta_{\rm atm}\simeq \theta_{23}$. 
The unstable ones are $\theta_{\odot}$ and 
$\Delta m_{\odot}^2 =m_2^2-m_1^2$. 
The angle $\theta_{13}\sim |V_{13}|$ and a Dirac phase $\delta$ are induced 
by the renormalization group. \\
As for $\Delta m_{\odot}^2$, we have
$\Delta m_{\odot}^2 =(\Delta_{21}-
4\epsilon m_1^2 s_{\rm atm}^2 \cos 2 \theta_{12})/{\cos 2\theta}$, 
where  $\tan 2\theta= 4\epsilon m_1^2 s_{\rm atm}^2
\sin 2 \theta_{12}\cos \frac{\a_0}2/(\Delta_{21}-
4\epsilon m_1^2 s_{\rm atm}^2 \cos 2 \theta_{12})$
In the following, we take the convention, $\Delta_{21}>0$ and 
$\Delta m_{\odot}^2 >0$ in accordance with 
$\tan^2 \theta_\odot =0.34 <1$. \\
As for the solar mixing angle, we have 
$s_{\odot}=| s_{12}c+ c_{12}s e^{i\a_0/2}|$, 
$c_{\odot}=| c_{12}c- s_{12}se^{-i\a_0/2}|$. The angle 
$s_{13}=\sin \theta_{13}$ is given by 
$|V_{13}|=\epsilon m_1m_3 \sin 2 \theta_{12}\times\sin 2\theta_{\rm atm}
\sin\frac{\alpha_0}{2}/\Delta m^2_{\rm{atm}}$. 
The Dirac phase, $\delta$ and two Majorana 
phases, $\a_M$ and $\b_M$ are given explicitly[6].
\\
(1) $\tan^2 \theta_\odot$: The solar mixing angle $\theta_\odot$ 
at $m_Z$ and $\theta_{12}$ at $M_R$ are related as
\bea
\hspace*{-1cm}
1+\cos\a_0
=\frac{-|\cos 2\theta_\odot|
-h\cos^22 \theta_{12}+ \cos2 \theta_{12}
\sqrt{h^2\cos^22 \theta_{12}+2|\cos 2\theta_\odot|h+1}
}{h\sin^22 \theta_{12}}\;,\nonumber\\
\ena
where $h=2\epsilon m_1^2 s_{\rm atm}^2/\Delta m^2_{\odot}$. 
From this formula, we see that 
the solar mixing angle, $\theta_\odot$ is given by 
$m_1$ (the mass at $m_z$), $\theta_{12}$ (the angle at $M_R$), 
and $\alpha_0$ (Majorana phase), if we use $\theta_{\rm atm}$, 
$\Delta m_{\odot}^2$ and $\epsilon$. 
We computed the relation among $m_1$, $\theta_{12}$ and $\alpha_0$ 
by taking $\tan^2 \theta_\odot=0.37$, 
$\Delta m_{\odot}^2=5\times 10^{-5}{\rm eV}^2$ and 
$\sin^2 2\theta_{\rm atm}=1$ for $\tan \beta=50$. The result 
is shown in Fig.1, where we observe that the wide range of 
$\theta_{12}$ (the solar angle at $M_R$) can  reproduce the 
experimental solar angle $\theta_\odot$ when for $m_1\sim 0.05$ eV and 
$\pi/2<\alpha_0<\pi$. 
By requiring $0\leq 1+\cos\alpha \leq 2$, 
\bea
\frac{\sin^2 2\theta_{\odot}}{\sin^2 2\theta_{\odot}
+(|\cos 2\theta_{\odot}|
+4\epsilon s_{\rm atm}^2 (m_1^2/\Delta m_{\odot}^2) )^2
}\leq \sin^22 \theta_{12}
\leq \sin^22\theta_{\odot}\;.
\ena
(2) $|V_{13}|$: This quantity is proportional to 
$\epsilon m_1m_3/\Delta m_{\rm atm}^2$ so that it is small.  
In Fig.2, we plot the predicted value of $|V_{13}|(m_1/m_3)$ in 
the $\sin^2 \theta_\odot$ and $m_1$ plane for $\tan \beta=50$. 
We observe that $|V_{13}|\sim 0.05$ may be expected when 
$m_1\sim 0.05$eV. 
\\
(3) $\delta$: We parametrize it as $\delta=\delta_1-\beta_0$. 
Then, we can predict $\delta_1$ as a function of $m_1$, 
$\theta_{12}$ and $\alpha_0$. In Fig.3, we show the expected 
value of $\sin \delta_1$ ($\delta_1=\delta-\beta_0$). 
\\
(4) $\langle m_{\nu}\rangle$: The effective mass 
for the neutrinoless double beta decay in this mode is given by
\bea
\langle m_{\nu}\rangle
\equiv \left | \sum_j m_j V_{ej}^2 \right | 
\simeq m_1\sqrt{1-\sin^22\theta_{12}\sin^2\frac{\alpha_0}{2}}
\sim m_1\;,
\ena
where we used $m_1\simeq m_2$. If $m_1\sim 0.05$eV, we expect 
$\langle m_{\nu}\rangle$ to be the same order as seen in Fig.3. \\
{\it Summary}: 
The assumption that $|V_{13}|=0$ at $M_R$ scale is attractive 
because this it enables to predict the value of $|V_{13}|$ and 
the Dirac phase is induced by the renormalization group, through 
two Majorana phases. Since at high energy there are only 
Majorana phases, which may be related to Majorana phases  
which contribute to Leptogenesis. 

\begin{figure}
\begin{center}
\includegraphics[scale=0.7]{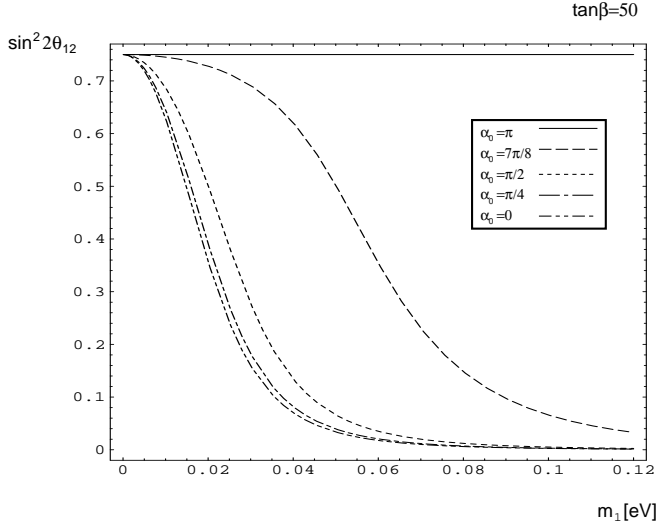}
\caption{\scriptsize Contour plot of $\alpha_0$ in
$\sin^2 2\theta_{12}$ and $m_1$ plain
to reconstruct the experimental value of
$\theta_{\odot}$ in the case of $\tan\beta=50$.
We use as experimental values
$\tan^2\theta_{\odot}=0.34$($\sin^22\theta_{\odot}\simeq 0.75$),
$\Delta m^2_{\odot}=5\times 10^{-5}[\mathrm{eV}^2]$,
$\sin^22\theta_{\mathrm{atm}}=1$, and
$\Delta m_{\mathrm{atm}}^2=3\times 10^{-3}[\mathrm{eV}^2]$.}
\end{center}
\end{figure}
\begin{figure}
\begin{center}
\includegraphics[scale=0.7]{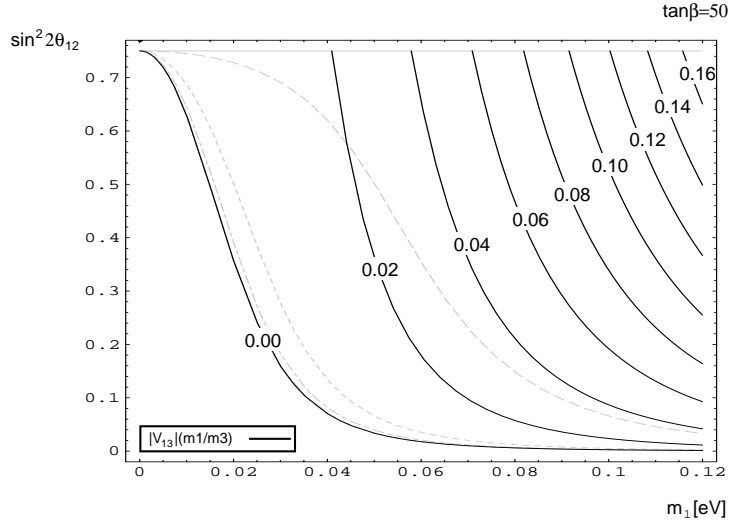}
\caption{\scriptsize 
 Contour plot of $|V_{13}|(m_1/m_3)$ in $\sin^22\theta_{12}$
and $m_1$ plain for $\tan\beta=50$.
We use same values as Fig.~1 for experimental values.
Gray curves show the $\alpha_0$ values as in Fig.~1.}
\end{center}
\end{figure}

Acknowledgment: 
This work is supported in part by 
the Japanese Grant-in-Aid for Scientific Research of
Ministry of Education,  Culture, Sports, Science and 
Technology (No.12047218).
\vskip 1mm

\end{document}